\documentclass[11pt]{article}
\usepackage{graphicx}
\usepackage{epsfig}
\newcommand{\BABARPubYear}    {00}

\newcommand{\BABARConfNumber} {07}
\newcommand{\SLACPubNumber} {8529}

\def\D    {\ensuremath{D}}
\def\Dp    {\ensuremath{D^+}}
\def\Dm    {\ensuremath{D^-}}
\def\Dstar    {\ensuremath{D^{*}}}
\def\Dstarp    {\ensuremath{D^{*+}}}
\def\Dstarm    {\ensuremath{D^{*-}}}

\def\rhop    {\ensuremath{\rho^+}}

\def\aonep    {\ensuremath{a_1^+}}

\usepackage{relsize}
\def\babar{\mbox{\slshape B\kern-0.1em{\smaller A}\kern-0.1em
    B\kern-0.1em{\smaller A\kern-0.2em R}}}

\def\epem       {\ensuremath{e^+e^-}}

\def\mumu       {\ensuremath{\mu^+\mu^-}}

\def\piz   {\ensuremath{\pi^0}}

\def\pip   {\ensuremath{\pi^+}}
\def\pim   {\ensuremath{\pi^-}}

\def\Kbar  {\kern 0.2em\overline{\kern -0.2em K}{}}

\def\Kp    {\ensuremath{K^+}}
\def\Km    {\ensuremath{K^-}}
\def\KS    {\ensuremath{K^0_{\scriptscriptstyle S}}} 
 
\def\Kstarz  {\ensuremath{K^{*0}}}

\def\Kzb   {\ensuremath{\Kbar^0}}
\def\KzKzb {\ensuremath{K^0 \kern -0.16em \Kzb}}
\def\Dz    {\ensuremath{D^0}}
\def\Dbar  {\kern 0.2em\overline{\kern -0.2em D}{}}

\def\Dzb   {\ensuremath{\Dbar^0}}
\def\DzDzb {\ensuremath{D^0 {\kern -0.16em \Dzb}}}
\def\Dstar   {\ensuremath{D^*}}

\def\Dstarz  {\ensuremath{D^{*0}}}
\def\Dstarzb  {\ensuremath{\Dbar^{*0}}}

\def\Bz    {\ensuremath{B^0}}
\def\B     {\ensuremath{B}}
\def\Bbar  {\kern 0.18em\overline{\kern -0.18em B}{}}

\def\Bzb   {\ensuremath{\Bbar^0}}
\def\Bu    {\ensuremath{B^+}}
\def\Bub   {\ensuremath{B^-}}
\def\Bpm   {\ensuremath{B^\pm}}

\def\BB    {\ensuremath{B\Bbar}} 
\def\BzBzb {\ensuremath{B^0 {\kern -0.16em \Bzb}}}

\def\jpsi  {\ensuremath{{J\mskip -3mu/\mskip -2mu\psi\mskip 2mu}}} 
\def\psitwos {\ensuremath{\psi{(2S)}}}
\mathchardef\Upsilon="7107
\def\Y#1S{\ensuremath{\Upsilon{(#1S)}}}% no space before {...}!

\def\FourS {\Y4S}

\mathchardef\Deltares="7101
\mathchardef\Xi="7104
\mathchardef\Lambda="7103
\mathchardef\Sigma="7106
\mathchardef\Omega="710A
\def\Deltabar   {\kern 0.25em\overline{\kern -0.25em \Deltares}{}}
\def\Lbar {\kern 0.2em\overline{\kern -0.2em\Lambda\kern 0.05em}\kern-0.05em{}}
\def\Sigbar{\kern 0.2em\overline{\kern -0.2em \Sigma}{}}
\def\Xibar{\kern 0.2em\overline{\kern -0.2em \Xi}{}}
\def\Obar{\kern 0.2em\overline{\kern -0.2em \Omega}{}}
\def\Nbar{\kern 0.2em\overline{\kern -0.2em N}{}}
\def\Xbar{\kern 0.2em\overline{\kern -0.2em X}{}}

\def\upsbb {\ensuremath{\Upsilon{\rm( 4S)}\to B\Bbar}}

\def\pt         {\mbox{$p_T$}}
\def\mes        {\mbox{$m_{\rm ES}$}}

\def\ev   {\ensuremath{\rm \,e\kern -0.08em V}}
\def\kev  {\ensuremath{\rm \,ke\kern -0.08em V}} 
\def\mev  {\ensuremath{\rm \,Me\kern -0.08em V}} 
\def\gev  {\ensuremath{\rm \,Ge\kern -0.08em V}} 
\def\gevc {\ensuremath{{\rm \,Ge\kern -0.08em V\!/}c}} 
\def\tev  {\ensuremath{\rm \,Te\kern -0.08em V}}
\def\mevc {\ensuremath{{\rm \,Me\kern -0.08em V\!/}c}} 
\def\gevcc{\ensuremath{{\rm \,Ge\kern -0.08em V\!/}c^2}} 
\def\mevcc{\ensuremath{{\rm \,Me\kern -0.08em V\!/}c^2}}

\def\cm   {\ensuremath{\rm \,cm}}

\def\mm   {\ensuremath{\rm \,mm}}

\def\mum  {\ensuremath{\,\mu\rm m}} %% mu meter 

\def\invfb   {\ensuremath{\mbox{\,fb}^{-1}}}
\def\mus  {\ensuremath{\rm \,\mus}}

\def\ps   {\ensuremath{\rm \,ps}}
\def\fs   {\ensuremath{\rm \,fs}}

\def\mus        {\ensuremath{\,\mu{\rm s}}}    %% microsecond
\def\ps         {\ensuremath{{\rm \,ps}}}   %% picosecond
\def\mrad{\ensuremath{\rm \,mr}}                %% milliradian

\def\gsim{{~\raise.15em\hbox{$>$}\kern-.85em
          \lower.35em\hbox{$\sim$}~}}
\def\lsim{{~\raise.15em\hbox{$<$}\kern-.85em
          \lower.35em\hbox{$\sim$}~}}

\def\CP                 {\ensuremath{C\!P}}
\def\ra                 {\ensuremath{\rightarrow}}
\def\to                 {\ensuremath{\rightarrow}}

\def\pep2{PEP-II}
\def\BF{$B$ Factory}
\def\abf {asymmetric \BF}

\newcommand{\dedx}{\ensuremath{\mathrm{d}\hspace{-0.1em}E/\mathrm{d}x}}

\newcommand{\eqref}[1]{Eq.~(\ref{eq:#1})}

\newcommand{\epjc}      [1]  {{Eur.\ Phys.\ Jour.\ C~{\bf #1}}}

\def\jetset74   {\mbox{\tt Jetset \hspace{-0.5em}7.\hspace{-0.2em}4}}
\def\B  {\ensuremath{B}}

\long\def\inst#1{\par\nobreak\kern 4pt\nobreak
    {\it #1}\par\vskip 10pt plus 3pt minus 3pt}

\begin{document}
{\thispagestyle{empty}

\begin{flushright}
\babar-CONF-\BABARPubYear/\BABARConfNumber \\
%\babar-PUB-\BABARPubYear/\BABARPubNumber \\
SLAC-PUB-\SLACPubNumber
\end{flushright}

%\vspace*{-11pt}
%\begin{flushleft}
%BaBar Analysis Document \#72, Version 04
%\end{flushleft}

\par\vskip 3cm

% Title of the paper
\begin{center}
\Large \bf
A measurement of the charged and neutral {\boldmath $B$} meson lifetimes
using fully reconstructed decays
%\boldmath \B\ lifetime measurements using fully reconstructed \B\ decays
%\footnote
%{Editors: J.Chauveau, Chih-hsiang Cheng,  D.Kirkby, F. Martinez-Vidal, 
%C. Roat and J. Stark.}
\end{center}
\bigskip

\begin{center}
\large The \babar\ Collaboration\\
\mbox{ }\\
\today
\end{center}
\bigskip \bigskip

% Abstract
\begin{center}
\large \bf Abstract \rm
\end{center}
Data collected with the
\babar\ detector at the \pep2\ asymmetric \BF\ at SLAC are used to
study the lifetimes of the \Bz\ and \Bu\ mesons.
The data sample consists of
7.4 \invfb\ collected near the \FourS\  resonance.
\Bz\ and \Bu\ mesons are fully reconstructed in several exclusive hadronic
decay modes to charm and charmonium final states.
The \B\ lifetimes are determined from the flight length difference between 
the two \B\ mesons which are pair-produced in the \FourS\ decay.
The preliminary measurements of the lifetimes are
\begin{eqnarray*}
\tau_{\Bz} &=& 1.506\pm 0.052\ {\rm (stat)} \pm 0.029\ {\rm (syst)}\ \ps, \\
\tau_{\Bu} &=& 1.602\pm 0.049\ {\rm (stat)} \pm 0.035\ {\rm (syst)}\ \ps
\end{eqnarray*}
and of their ratio is
$$ \tau_{\Bu }/\tau_{\Bz }=1.065\pm 0.044 \ {\rm (stat)} \pm 0.021 \ {\rm(syst)}.$$
%\centerline{where the first error is statistical and the second systematic.}

\vfill

\begin{center}
Submitted to the XXX$^{th}$ International 
Conference on High Energy Physics, Osaka, Japan.
\end{center}

\newpage
}

% Input author list file
\begin{center}
\small

The \babar\ Collaboration
\bigskip

%% author list as of 24-Jul-2000 (580 authors) % hand edited removal of street addresses by dbmacf
%% 8-4-00: fixed addresses for UC campuses; added Sh. Rahatlou to UCSD: dbmacf
B.~Aubert,
A.~Boucham,
D.~Boutigny,
I.~De Bonis,
J.~Favier,
J.-M.~Gaillard,
F.~Galeazzi,
A.~Jeremie,
Y.~Karyotakis,
J.~P.~Lees,
P.~Robbe,
V.~Tisserand,
K.~Zachariadou
\inst{Lab de Phys.\ des Particules, F-74941 Annecy-le-Vieux, CEDEX, France}
% ...

\end{center}\newpage

% reset footnote counter
\setcounter{footnote}{0}

\section{Introduction}

Precise measurements of the charged and neutral \B\ meson lifetimes
are needed for
tests of theoretical models of heavy quark decays.
An uncertainty of around 1\% will be required
to distinguish between models~\cite{bib:Bigi,bib:Neubert} predicting 
different lifetime hierarchies among the \B\  hadrons.
The lifetimes are also
required for other measurements and they provide a check of our
understanding of the detector.
The current world averages~\cite{bib:PDG2000} for the \B\
lifetimes are
$\tau_{\Bz }=1.548\pm0.032$\ps\ and $\tau_{\Bu }=1.653\pm0.028$\ps\ and
for their ratio is
$\tau_{\Bu }/\tau_{\Bz }=1.062\pm0.029$.
%reach a relative precision of about 2\% and 3\% respectively.

This paper presents preliminary measurements of the \Bpm\ and
\Bz/\Bzb\ lifetimes
and their ratio performed with data collected
by the \babar\ detector at the \pep2\ asymmetric \BF.
At the \FourS, \B\ mesons are produced in \Bu\Bub\ and
\BzBzb\ pairs. In this analysis, one of the \B\ mesons is fully 
reconstructed in a  variety of clean hadronic 
two-body decay modes to charm and charmonium final states.
An inclusive technique is used to reconstruct the decay vertex of the
second \B\ in an event, and the lifetimes are determined from the
distance between the decay vertices of the two \B\ mesons. This novel
method, developed for use at an \abf, deals with event topologies 
similar to those addressed in the analyses of \CP\ asymmetries where
the time dependence is obtained from the distance between the two vertices.  
The measurement of \B~lifetimes using this technique is
therefore a validation of the \CP\ analyses.
The systematic 
errors are quite different from previous measurements of the lifetimes
and we expect substantial increases in the statistical precision
in the future.

The paper is organized as follows. 
The method used to determine the \B~meson lifetimes is outlined in
Section~\ref{sec:method}. The data sample and the detector components 
most relevant for
this measurement are reviewed in Section~\ref{sec:detector}. The event
selection criteria and sample composition
are described next in Section~\ref{sec:breco}. The determination
of the distance between the decay vertices of the two \B~mesons in a selected
event is discussed in Section~\ref{sec:vertex} together with the
corresponding resolution function. 
%They are necessary to disentangle the effects of the \B~
%lifetimes and the experimental resolution. 
The fitting procedure used
to extract the \B\ lifetimes from the neutral and charged \B~meson
samples is described in Section~\ref{sec:lifetimefit}.
Systematic uncertainties are discussed in
Section~\ref{sec:systematics}. The results are given at the end of the paper.

\section{The decay length difference technique}
\label{sec:method}
The \FourS\ decays exclusively into pairs of charged or neutral \B\ mesons.
At the \pep2\ \abf\ the \FourS\ decay products are Lorentz boosted 
($\beta \gamma \approx 0.56$)\footnote{The average boost of the 
center of mass frame is measured on a run-by-run basis using two-prong
events~\cite{bib:babar}.} and travel far
enough for their flight paths ($\approx 260$\mum) to be comparable to the experimental
resolution ($\approx 150$\mum).
%, average value of core and tailresolutions). 
Since no charged stable particles emerge from the 
\FourS\ decay point, the production point of the \B\ mesons is unknown and the
\B\ lifetimes have to be determined solely from their decay points.
The difference between the decay lengths of the two \B\ mesons in an
event has been used for that purpose in this analysis.

\begin{figure}[htbp]
\begin{center}\mbox{\psfig{file=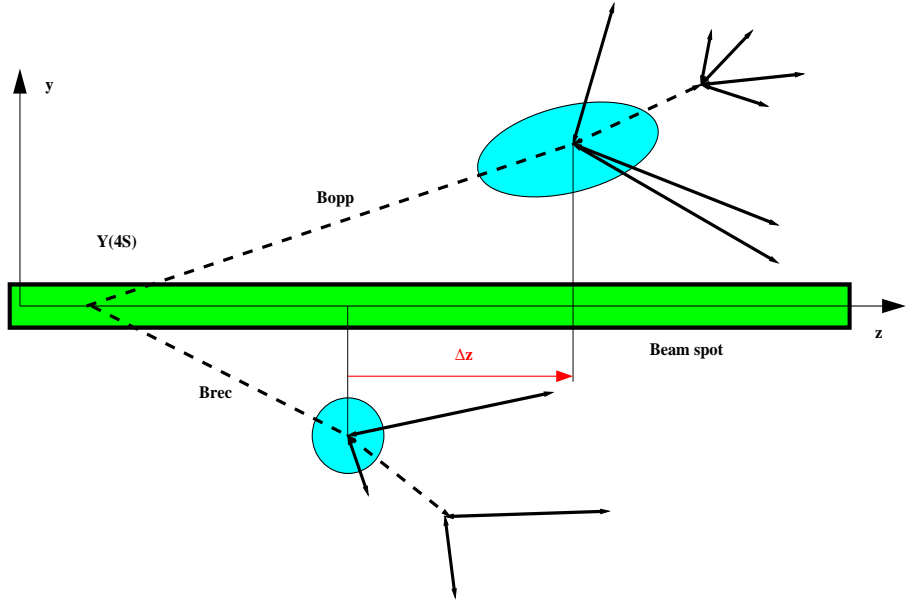,height=8cm,width=12cm
%,angle=-90
}
}
\end{center}
\caption{\label{fig:eventopo}Event topology. The figure is not drawn
to scale (see text).}
\end{figure}

Throughout the paper the signed quantity
$\Delta z = z_{\rm rec}-z_{\rm opp}$
is used for the projection on the $z$ axis (defined below) 
of the decay length difference between the fully 
reconstructed \B\ meson, thereafter called \B$_{\rm rec}$, and the
opposed \B\ meson, \B$_{\rm opp}$.
The event topology is sketched in Fig.~\ref{fig:eventopo},
which is not drawn to scale.
In the simplified situation where the \B\ mesons are produced
exactly at threshold and the \FourS\ moves exactly along the $z$ axis, 
$|\Delta z|$ is distributed exponentially with an average of
$\langle|\Delta z|\rangle = (\beta\gamma)_B c \tau_B =
(p_{\FourS}/m_{\FourS}) c \tau_B$.
Complications arise from several effects. The axis
of the \pep2\ beams is tilted by $20$\mrad\ with respect to the
$z$~axis, which is defined as parallel to the solenoidal magnetic
field, and the energies of the beams fluctuate, giving the 
\FourS\ momentum a Gaussian distribution with a standard 
deviation of 6\mevc. Furthermore, the energy
release in the \upsbb\ decay makes the \B\ mesons move in the \FourS\ rest
frame, resulting in a non-vanishing opening angle (smaller than 
$214$\mrad~\cite{bib:babbook113})
between the trajectories of the \B\ mesons.
The latter effect has the highest impact on the $\Delta
z$~distribution. In practice, all of these effects are small
compared to the experimental resolution on $\Delta z$.

%The flight paths are of the order of
%$260\ \mum $ and the errors on $z_{\rm rec}$ and $z_{\rm opp}$ are
%usually of the order of $50\ \mum $ and $150\ \mum $, respectively.

\section{The \babar~detector and data set}
\label{sec:detector}

The data used in this analysis were collected by the \babar\ 
detector
at the \pep2\ storage ring in the period  from January to June, 2000.
The total integrated luminosity of the data set is 7.4\invfb\
collected near the \FourS\ resonance and
0.9\invfb\ collected 40\mev\ below the resonance.
The corresponding number
of produced \BB\ pairs is estimated to be $8.4 \times 10^6$.

The \babar\ detector is described elsewhere~\cite{bib:babar}. For
this measurement, the most important subdetectors are the silicon
vertex tracker (SVT) and the central drift chamber (DCH), which jointly
provide charged particle tracking, and the CsI~electromagnetic calorimeter 
(EMC), from which photons and $\pi^0$s are reconstructed.
The charged particle transverse momentum (\pt) resolution is approximately 
described by the formula
$\left( \delta \pt / \pt \right)^2 =  (0.0015\, \pt)^2 + (0.005)^2$, where 
\pt\ is in \gevc.  The SVT, 
with a typical single-hit resolution of 10\mum, 
provides vertex information in both  
the transverse plane and in $z$. 
The precision on charged particle momenta, neutral particle energies
and spatial coordinates from the tracking and calorimetry 
leads to resolutions on invariant masses and other
kinematical quantities which are adequate for a
clean separation of the exclusive hadronic modes as described
in Section~\ref{sec:breco}.
Impact parameter resolutions in the
transverse and longitudinal coordinates are $\simeq 50$\mum\ at high momentum, and better
than 100\mum\ for $\pt > 0.6$\gevc.
The achieved precision on exclusive and inclusive vertices are given
in Section~\ref{sec:vertex}.

Particle identification is needed for complex decay modes and 
\jpsi\ reconstruction.
Leptons and hadrons are identified using a combination of measurements
from all \babar\ components, including 
the \dedx\ energy loss using a truncated mean of 
40 samples (maximum) in the DCH and 5 samples in the SVT.
Electrons and photons are identified in the barrel and forward regions 
by the EMC. Muons are identified in the
instrumented flux return (IFR).   In the central polar 
angle region the Cherenkov ring
imaging detector (DIRC)
%~\cite{bib:joe} 
provides a separation better than three
standard deviations between pions and kaons  over the upper part of the  
momentum range of \B\ decay products (above 700\mevc) where the
\dedx\ sensitivity vanishes.

%%%%%% ce que j'ai redige pour le papier conf %%%%%%
\section{Event selection}
\label{sec:breco}

\Bz\ and \Bu\ mesons are reconstructed in the following hadronic 
modes\footnote{Throughout this paper, when a mode is quoted, its
charge conjugate is implied.};
$\Bz \to D^{(*)-} \pi^+$, $D^{(*)-} \rho^+$,
$D^{(*)-} a_1^+$, $\jpsi \Kstarz$ and
$\Bu \to \overline{D}^{(*)0} \pip$,
$\jpsi K^+$, $\psitwos K^+$.
All final state particles are reconstructed.
The selection criteria are devised to obtain final samples with
approximately 90\% purity and have
been studied using the Monte Carlo simulation.
Optimization for background rejection was performed
using continuum data and sideband regions that do not contain signal events.
The details of the selection of these events are described
below. 

\subsection{Selection criteria}
The basic objects used in the selection are charged tracks and
neutral electromagnetic clusters. Charged tracks are built out of hits in 
the SVT and the DCH by a Kalman filter-based tracking 
algorithm~\cite{bib:babar}.
To be retained, a track must not miss the average beamspot position by
more than 1.5\cm\ in $xy$ and 10\cm\ in $z$.
Most tracks must penetrate halfway radially into the drift
chamber (at least 20 hits are required) and  have a momentum between
0.1 and 10\gevc. The latter restrictions do not apply to soft pions
or low momentum tracks from a high multiplicity
$D$~decay. Electromagnetic clusters in the EMC are photon candidates
when they carry more than 30\mev, have a narrow transverse shower
profile (the lateral moment defined in Ref.~\cite{bib:babar}) less than~0.8
and are not matched to a charged track.
 
The selection procedure begins by enforcing standard
criteria~\cite{bib:babar} on the charged multiplicity, total energy,
global event vertex position and event shape which enrich the
sample with multihadronic events from \FourS\ decays.
A tight requirement on the normalized second
Fox-Wolfram moment $R_2(=H_2/H_0) < 0.5$
is applied in all analyses pertaining to this paper
to reject jet-like events from the continuum.

The selection then proceeds by building
a tree of composite objects (fast decaying particles) from tracks and photon 
candidates, culminating in
a \B\ candidate if one exists in the event.
Geometrical fits are used to reconstruct
decay points and declared successful when the $\chi^2$ probability is
above~0.1\%. 
No goodness of fit requirement is enforced for the kinematic fits
used to impose mass constraints.

Neutral pion candidates are formed from pairs of photon candidates
with an invariant mass within 20\mevcc\ of the nominal \piz\ mass
and with a minimum energy of 200\mev, except for those used in attempts to
reconstruct a \Dstarz. A kinematic fit is performed
on the selected candidates imposing the \piz\ mass constraint.

\KS\ candidates are reconstructed in the two charged pion decay mode only.
Oppositely charged tracks are vertexed using a geometrical fit.
The \pip \pim\  invariant mass, computed at the vertex of the two
tracks, has to lie between 462 and 534\mevcc.
The opening angle between the
flight direction and the  \KS\ candidate momentum vector 
must be smaller than 200 \mrad. Finally, the transverse flight distance
from the event primary vertex is required to be greater than 2\mm.

$K$ and $\pi$ meson candidates are combined to search for the following
$D$ meson decay modes;
$\Dz\ra\Km \pip$, $\Km \pip \piz$, $\Km \pip \pip \pim$, $\KS \pip \pim$
and $\Dp\ra\Km \pip \pip$, $\KS \pip$.
All kaons and the pions from two body modes are required to have a momentum
above 200\mevc. This criterion is relaxed to 150\mevc\ for pions from three
or four body final states. Candidates with an invariant mass 
within three standard deviations of the known mass are retained. The
standard deviation computed event by event by the tracking algorithm 
was used after it was verified that the normalized error
distribution (pull) was correctly estimated at the 10-20\% level.
The reconstruction of $\Dz \to \Km \pip \piz$
is restricted to the dominant resonant
mode $\Dz \to \Km \rhop$, $\rhop \to \pip \piz$. 
The \pip \piz\ mass is required to lie within $\pm 150$\mevcc\ of the 
known $\rho$ mass and the $\rho$ decay helicity angle in its rest frame,
$\theta^*_\rho$, is required to satisfy
$|\cos \theta^*_\rho| > 0.4$. All $D$ candidates must have a 
momentum greater than 1.3\gevc\ in the $\FourS$ frame and have a
successful vertex fit. The accepted $D$ candidates are 
mass constrained before they are used in subsequent intermediate state 
searches. Charged (neutral) $D^*$ mesons are built by combining a \Dz\ with 
a soft charged (neutral) pion with momentum above 70 (100)\mevc. 
Constraining the vertex of the \Dstarp\ to the beamspot vertical position
within 40\mum\ improves the resolution on the soft pion direction
and hence on $\Delta m$, the difference between the \Dstar\ and the
$D$ candidate invariant masses.
This procedure has been verified to be bias-free
in Monte Carlo events.
An acceptable range of $\pm$3~standard deviations around the
nominal $\Delta m$ completes the \Dstar\ selection. The range limits
($\pm 1.1$\mevcc\ for $\Dz \to \Km \rhop$, $\pm 0.8$\mevcc\ for all other
$D^{*+}$ modes and $\pm 1.4$\mevcc\ for the \Dstarz)
are weighted averages of the standard deviations obtained from
two-Gaussian fits to the  experimental $\Delta m$ distributions.

The charmonium meson selection is the same as that described 
in Ref.~\cite{bib:BabarPub0005}.
Leptonic decays of charmonium mesons are reconstructed from pairs of
oppositely charged tracks considered either as electron or muon
pairs. At least one decay product must be positively identified as a
lepton. For an electron inside the EMC acceptance, this means a positive
electromagnetic shower signature, otherwise that 
the measured \dedx\ be consistent with the electron hypothesis. For muon pairs,
the second track has to be minimum ionizing in the EMC. In the
electron case, when both tracks are inside the EMC acceptance and are
such that $E/p>0.5$, where $E$ is the energy measured in the 
EMC and $p$ is the momentum obtained from the tracking, bremsstrahlung photons
found in the EMC are combined with the tracks to reconstruct the
electrons prior to radiation. $\jpsi \ra \epem$ (\mumu)
candidates must have an invariant mass between 
2.95 (3.06) and 3.14\mevcc. The  invariant mass of the 
$\psitwos \to \mumu$ candidates is required to be within 50\mevcc\ 
of the known $\psitwos$ mass, but a less stringent lower limit of
200\mevcc\ is imposed for $\psitwos \to \epem$ candidates. 
$\psitwos \ra \jpsi \pip \pim$  decays are also reconstructed
by combining  oppositely charged pions with mass-constrained
\jpsi\ candidates and selecting the \psitwos\ invariant 
mass between 3.1 and 4.0\gevcc\ and the $\psitwos -\jpsi$ mass
difference in a window of $\pm 15$\mevcc\ around the nominal value.

\B\ candidates are formed by combining a \Dstar, \D, \jpsi~or
\psitwos\ candidate with a \pip, \rhop, \aonep, \Kstarz\ or
\Kp\ candidate having  a minimum momentum of 500\mevc\ in the
\FourS\ frame. For $\Bz \to D^{(*)-} \rhop$, the \piz\ meson from 
the \rhop\ decay is required to have an
energy higher than 300\mev. For $\Bz\to  D^{(*)-} \aonep$,
the \aonep\ meson, selected by combining three charged
pions, must have an invariant mass between 1.0 and 1.6\mevcc\ and
a successful vertex fit. 

Kaon identification is used to
reject background at the price of some loss in reconstruction efficiency.
For most of the \Bz\ modes, a loose kaon selection, or no selection,
is enough to reduce the background level to the required level. 
Tighter kaon identification is mandatory only for modes with high combinatorial
background, such as $\Bz \to \Dm \aonep$.

Finally, to achieve the required high signal to background ratio,
stricter requirements on  event shape variables are applied.
A selection is performed on the angle, $\theta_T$, between two thrust axes,
namely the thrust axis computed from the
tracks of the $B$~meson candidate and the thrust axis of all the other
charged tracks and neutral clusters in the event. A flat distribution is
expected for \B\ meson pairs while continuum events tend to peak
at $\cos\theta_T = \pm 1$. 
For the $\B \to \Dstar \rho$ modes,
$|\cos\theta_T|$ is required to be less than 0.9 
for the $\Dz \to K^-\pip$ and $K^-\pip \piz$ channels and less 
than 0.8 for $\Dz \to K^- \pip \pip \pim$ and $\KS \pip \pim$. The
requirements are 0.8 and 0.7 for the $\Bz\to \Dstarm \aonep$ modes 
and the same two couples of $D$ modes. There is no restriction based on
thrust for the $\B \to D^{*-}\pip$ decays.
For the \Bz\ decaying to charged $D$ mesons,
the requirements are 0.9, 0.8 and 0.7, if the \Dm\ is accompanied by a 
\pip, a \rhop\ or an \aonep. For the $\Bub \to D^{(*)0}\pim$ decays, 
$|\cos\theta_T|$ is required to be less than 0.9 in the case
of the $\Dz \to K^-\pip$ decay and to be less than 0.8 in all other cases.
The requirement is 0.9 for the \B\ decays to charmonium.

\subsection{Sample composition}
\label{subsec:sample}
Signals for each decay mode in the selected sample are isolated by
analyzing the two-dimensional distribution of the
kinematical variables, $\Delta E$ and \mes, which have an
intuitive meaning in the \FourS\ rest frame. $\Delta E=E^*_{\rm rec}-E^*_b$
is the difference of the \B\ candidate energy and the beam
energy. $\mes=\sqrt{E^{*2}_b - \mbox{\boldmath $p$}^{*2}_{\rm rec}}$ is the
mass of a particle with a reconstructed momentum 
$\mbox{\boldmath $p$}^*_{\rm rec} = \sum_i \mbox{\boldmath $p$}^*_i$
assumed to have the beam energy, as is the case for a true \B\ meson. 
Candidates with $ 5.2 < \mes < 5.3$\mevcc\ and
$\left| {\rm \Delta} E \right| < 140$\mev\ are considered in
further analyses. In case an event has several \B\ candidates, only the one
with the smallest $\Delta E$ is kept.
The $\Delta E$ and \mes\ variables are essentially uncorrelated. The
resolution on \mes\ is about 3\mevcc\ and is dominated by 
the energy spread of the beams,
inherent to the collider. The resolution on
$\Delta E$ varies from mode to mode between 12 and 40\mev.  
For each mode a rectangular signal region is defined by the three
standard deviation bands in \mes\ ($5.27 < \mes < 5.29$\gevcc)
and $\Delta E$ (an interval which depends on the mode).  

\begin{figure}[tbhp]
\begin{center}
\begin{tabular}{lr}
\mbox{\epsfxsize=8cm\epsffile{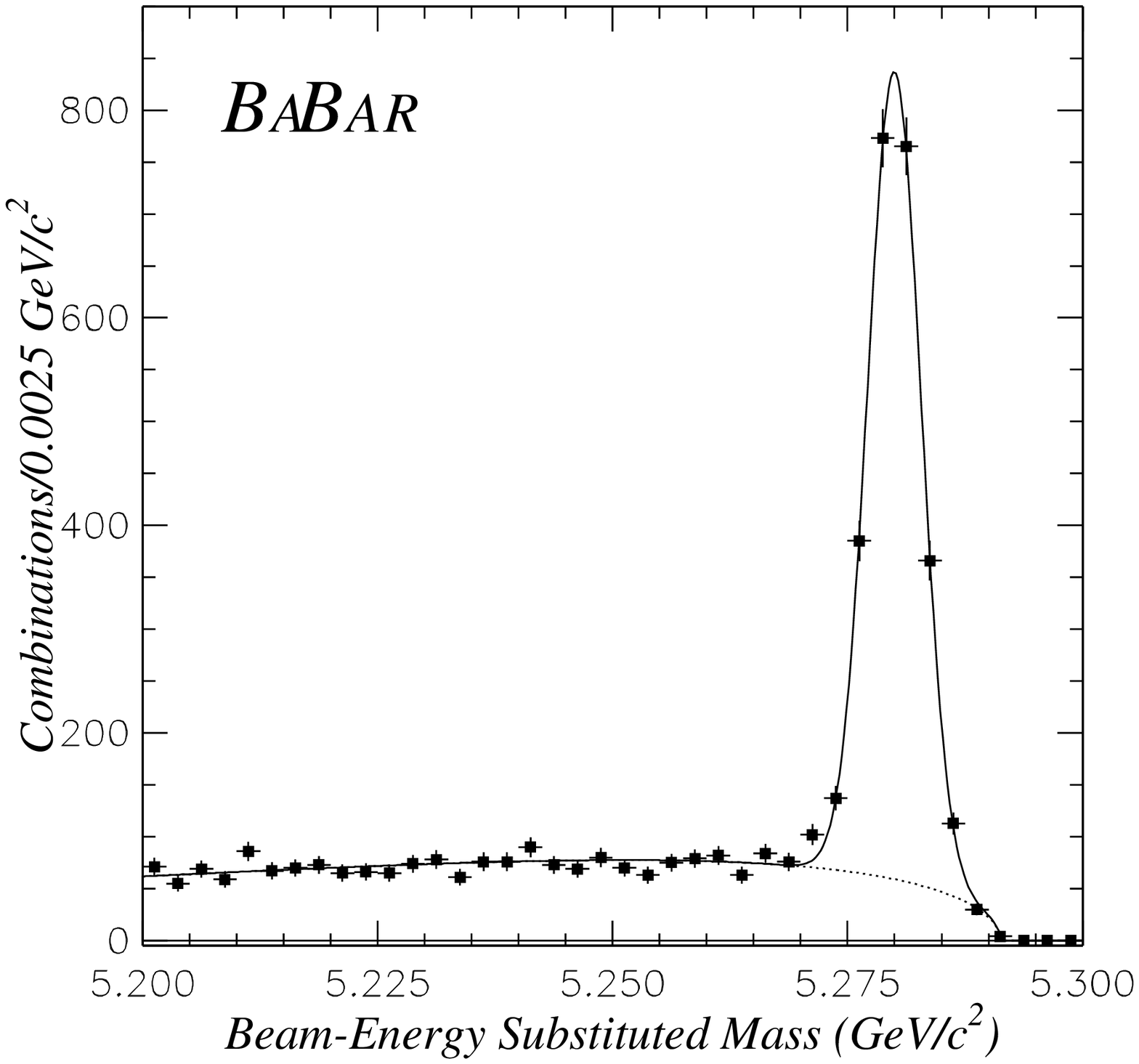}} &
\mbox{\epsfxsize=8cm\epsffile{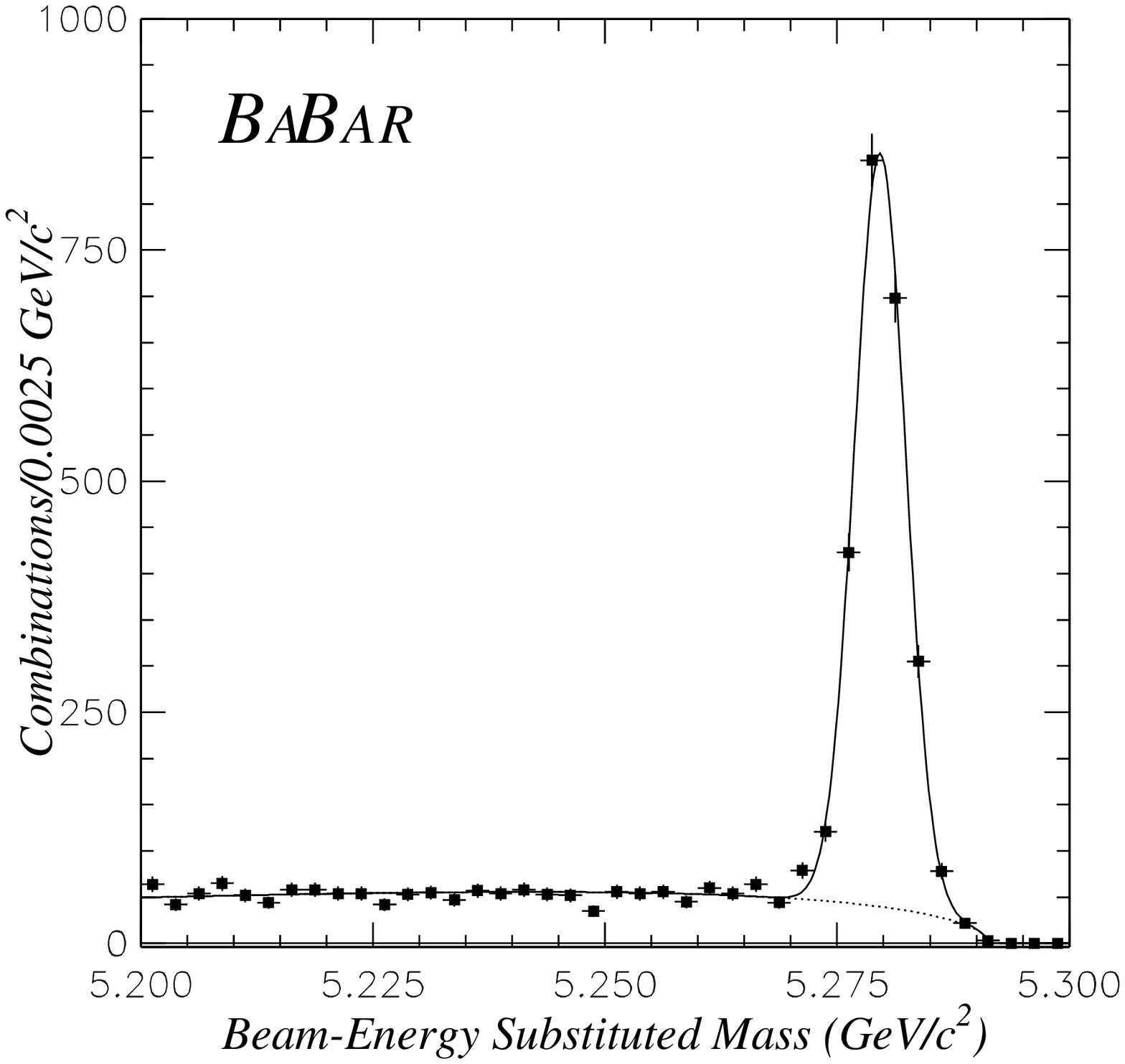}} \\
\end{tabular}
\vskip-6.7cm~~~~~~~~~~~~~~~~~~~~~~~~~~~~~~~~~~~~~~~~~~~~~{\Large
  (a)}~~~~~~~~~~~~~~~~~~~~~~~~~~~~~~~~~~~~~~~~~~~~~~~~~~~~~~~~~~~~~{\Large
  (b)}\\
\vskip6.2cm 
\end{center}
\caption{\mes\ distribution for all the hadronic modes for (a) \Bz\ and 
(b) \Bu. The signal is fitted to a Gaussian and for background we
use the ARGUS parameterization \cite{bib:argusfunction}. The total
numbers of signal events in all \Bz\ and \Bu\ modes are $2210\pm58$ 
and $2261\pm53$ respectively.
\label{fig:hadronicb0bch}}
\end{figure}

For each mode, the sample composition was determined by fitting the
\mes\ distribution, for candidates within the
signal region in $\Delta E$,
to the sum of a single Gaussian representing the signal
and a background function used by the ARGUS
collaboration~\cite{bib:argusfunction}. The purity of each subsample
is computed as the ratio of the area of the Gaussian in the $\pm 3
\sigma$ range over the total area in this range. The results are summarized in 
Table~\ref{tab:HadronicBYield}. Figure~\ref{fig:hadronicb0bch}
shows the \mes\ distributions for the summed  hadronic
\Bz\ and \Bu\ modes with the fits superimposed.

To extract separately the \Bz\ and \Bu\ lifetimes, a
thorough understanding of the background shape and composition is necessary.
In the signal region, the background receives 
contributions from continuum, \BzBzb\ and $B^+ B^-$ pairs in various
proportions for the different modes. A consistent background shape is
found from studies of several samples; namely, on-resonance data events
outside the signal region, off-resonance data, and generic
\BB\ and continuum Monte Carlo samples.
The contamination of the \Bu\Bub\ subsamples by \BzBzb\ events,
and vice versa, is found to be small.

%It was proven 
%that the contamination in the subsamples associated with charged
%(resp. neutral) \B 's by
%events coming fom neutral (resp. charged) \B\ mesons was small. More
%details are given below in section~\ref{sec:lifetimefit}. 
%The well understood properties of the  events from the $\Delta E$ sidebands
%in the \FourS\ data were used to determine by extrapolation
%the shape of the decay distance $\Delta z$ distribution of the
%background in the \B meson samples.  

\begin{table}[!htb]
\caption{Two-body hadronic \Bz\ and \Bu\ decay candidate yields and signal
purities from the fit to the \mes\ distribution. Signal purities are
estimated for $\mes > 5.27$\gevcc.}
\vspace{0.3cm}
\begin{center}
\begin{tabular}{|l|c|c|} \hline
  Decay mode   & Number of \B\ candidates  & Purity (\%) \\
\hline  \hline
 \Bz~$\to$ \Dstarm \pip       &  \phantom{1}552$\pm$26  & 90  \\ 
 \Bz~$\to$ \Dstarm \rhop      &  \phantom{1}374$\pm$23  & 84  \\ 
 \Bz~$\to$ \Dstarm \aonep     &  \phantom{1}202$\pm$18  & 79 \\ 
 \Bz~$\to$ \Dm \pip           &  \phantom{1}537$\pm$25  & 90 \\  
 \Bz~$\to$ \Dm \rhop          &  \phantom{1}279$\pm$20  & 84  \\ 
 \Bz~$\to$ \Dm \aonep         &  \phantom{1}194$\pm$18  & 73  \\ 
 \Bz~$\to$ \jpsi \Kstarz      &  \phantom{1}167$\pm$15  & 90 \\
 \Bu~$\to$ \Dzb \pip          &            1528$\pm$43  & 88 \\  
 \Bu~$\to$ \Dstarzb \pip      &  \phantom{1}446$\pm$25  & 89 \\  
 \Bu~$\to$ \jpsi,\psitwos \Kp &  \phantom{1}294$\pm$17  & 99 \\
\hline\hline
  Total \Bz & 2210$\pm$58  & 86  \\ 
  Total \Bu & 2261$\pm$53  & 89  \\ 
\hline
\end{tabular}
\end{center}
\label{tab:HadronicBYield}
\end{table}

\section{Vertex reconstruction and resolution function}
\label{sec:vertex}
To extract the \B\ meson lifetimes from the sample, the 
separate decay vertices have to be identified
and the distance between them along the $z$ axis, $\Delta z$, evaluated. 
The resolution function for $\Delta z$ is the most
critical ingredient of this analysis as the width of the $\Delta z$
distribution is due to the combined effect of the \B\ lifetime and the
detector resolution.

A geometric and kinematic fit of the fully
reconstructed~$\B_{\rm rec}$ is performed and is required  to converge. The
masses of $D$~and \jpsi\ resonances are fixed at the
nominal values~\cite{bib:PDG2000}. Furthermore, the vertices of short-living
resonances are constrained to be identical to the \B\ vertex.
Two-Gaussian fits
adequately describe the error distributions of the full Monte Carlo 
simulation used to
model the signal events, with standard deviations 
ranging from 45 to 65\mum\ and normalized fluctuations (called pulls
in the rest of the paper)  around~1.1.

The vertex of the other \B\ is determined via an inclusive and
iterative procedure applied to all the tracks that are not associated
with the $B_{\rm rec}$. To avoid biases from long-lived particles such as
neutral strange particles decaying to two charged prongs (a topology
known as a V$^0$ decay), as many
V$^0$ decays as possible are reconstructed and subsequently treated as single
tracks. A common vertex is then fitted using all tracks including the
$\B_{\rm opp}$ ``pseudotrack'' which is
obtained from the \FourS\ decay kinematics
using the $\B_{\rm rec}$ and the beamspot. The procedure is
repeated, after removing tracks and V$^0$ decays which result in poor fits,
until stability is achieved.
The algorithm is highly efficient and successful in
minimizing  biases due to the secondary charm decays of the $B_{\rm opp}$.
From the weighed means of two-Gaussian fits to Monte Carlo 
distributions of the residuals, one standard deviation errors of the
order of 115\mum\ and biases around 25\mum\ are obtained. Although
still present, the bias due to the charm particle lifetime is
kept small.

\begin{figure}[htbp]
\begin{center}
\begin{tabular}[t]{cc}
%\mbox{\psfig{file=Figures/vtxtag1.eps,height=8cm,width=8cm}} &
%\mbox{\psfig{file=Figures/vtxtag2.eps,height=8cm,width=8cm}} \\
%\mbox{\psfig{file=Figures/vtxtag3.eps,height=6cm,width=6.5cm}} &
%\mbox{\psfig{file=Figures/vtxtag4.eps,height=6cm,width=6.5cm}}
\mbox{\psfig{file=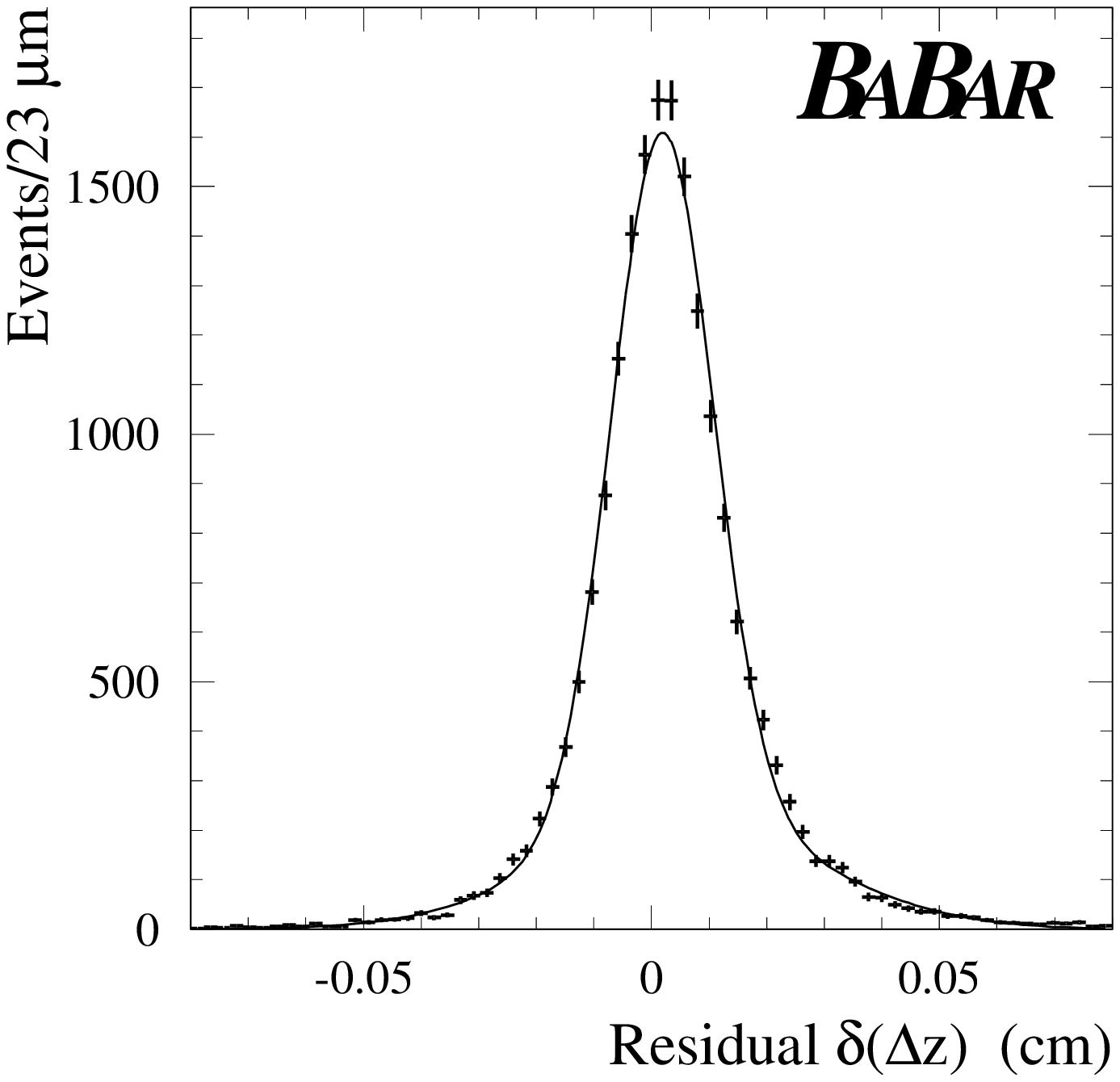,width=8.0cm}} &
\mbox{\psfig{file=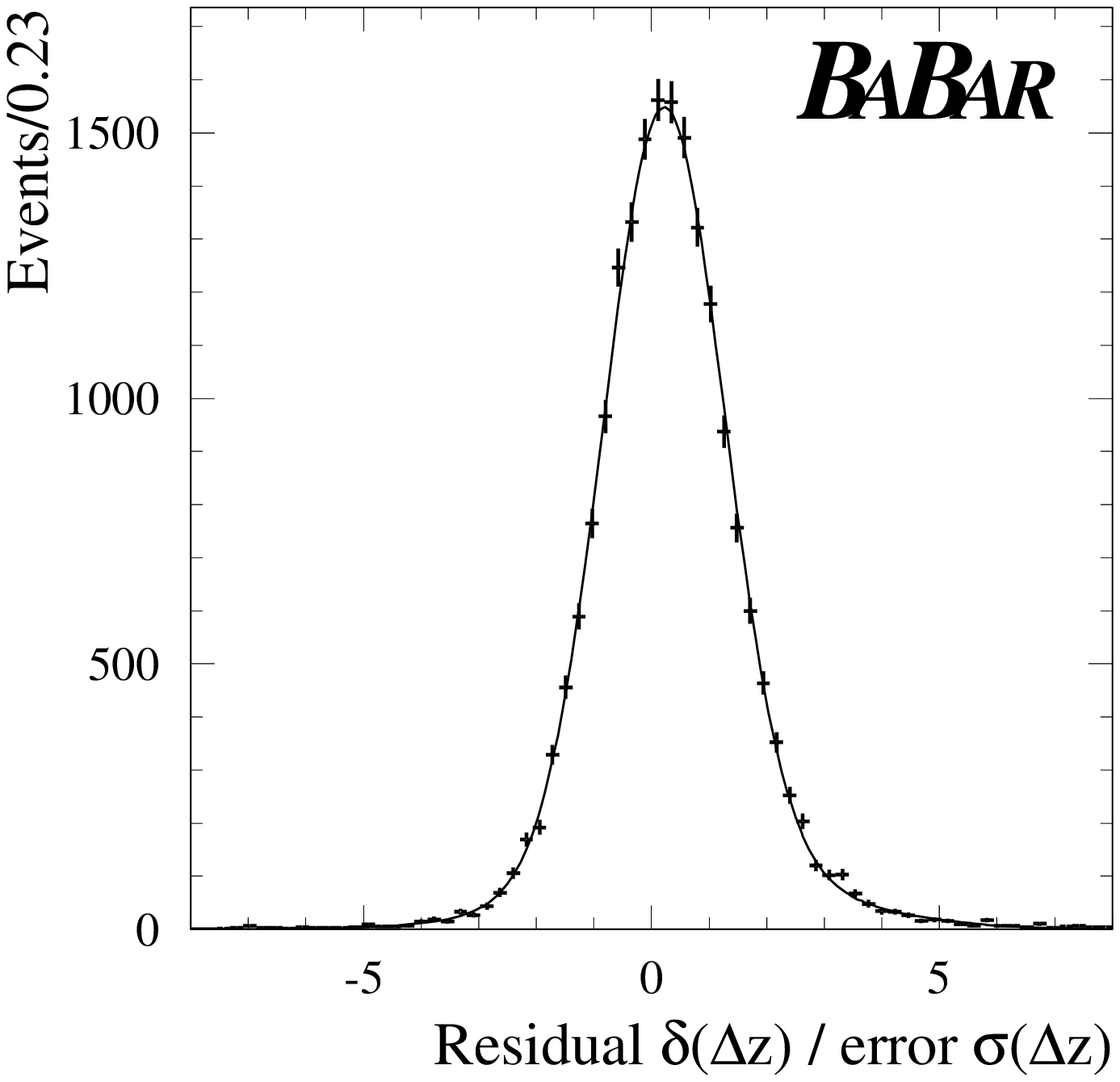,width=8.0cm}}
\end{tabular}
\vskip-6.9cm{\Large
  (a)}~~~~~~~~~~~~~~~~~~~~~~~~~~~~~~~~~~~~~~~~~~~~~~~~~~~~~~~~~~~~{\Large
  (b)}~~~~~~~~~~~~~~~~~~~~\\
\vskip6.4cm
\end{center}
\caption{
\label{fig:vtxtagdz}
Distribution of (a) residual and (b) pull of $\Delta z$ from a sample of
$\Bu \ra \jpsi \Kp$ Monte Carlo simulated events.
}
\end{figure}

With both vertices reconstructed, the decay length difference $\Delta z$ is
obtained. The Monte Carlo simulation
was used to study the $\Delta z$ resolution
function $\cal R$ as a function of two variables, the residual 
$\delta(\Delta z)= (\Delta z)_{\rm rec}-(\Delta z)_{\rm generated}$
and the pull $\delta(\Delta z)/\sigma(\Delta z)$.
Figure~\ref{fig:vtxtagdz} presents the distributions for the decay
$\Bu \ra \jpsi \Kp$.
Two-Gaussian fits successfully describe the
Monte Carlo residual and pull distributions. They give 
a one standard deviation width
of $130 \pm 1$\mum\ and a bias of $24.5 \pm 1$\mum\ for the
residual. The corresponding values for the pull are $1.21 \pm 0.01$ and
$0.29 \pm 0.01$ respectively. 
As expected their widths are dominated by the $B_{\rm opp}$ vertex errors.
Tails remain due to charm decays of the $B_{\rm opp}$. Because it is
governed by the $B_{\rm opp}$
vertex, the resolution function shape is the
same for all modes and also for neutral and charged \B\ mesons.
These comparisons are summarized in Table~\ref{tab:vtxtagmulti}.
%(Figure~\ref{fig:vtxtagmulti})

\begin{figure}[htbp]
\begin{center}
%\begin{tabular}
%[t]{ccc}
%\mbox{\psfig{file=Figures/fitparam3.eps,height=5cm,width=5cm}} &
\mbox{\psfig{file=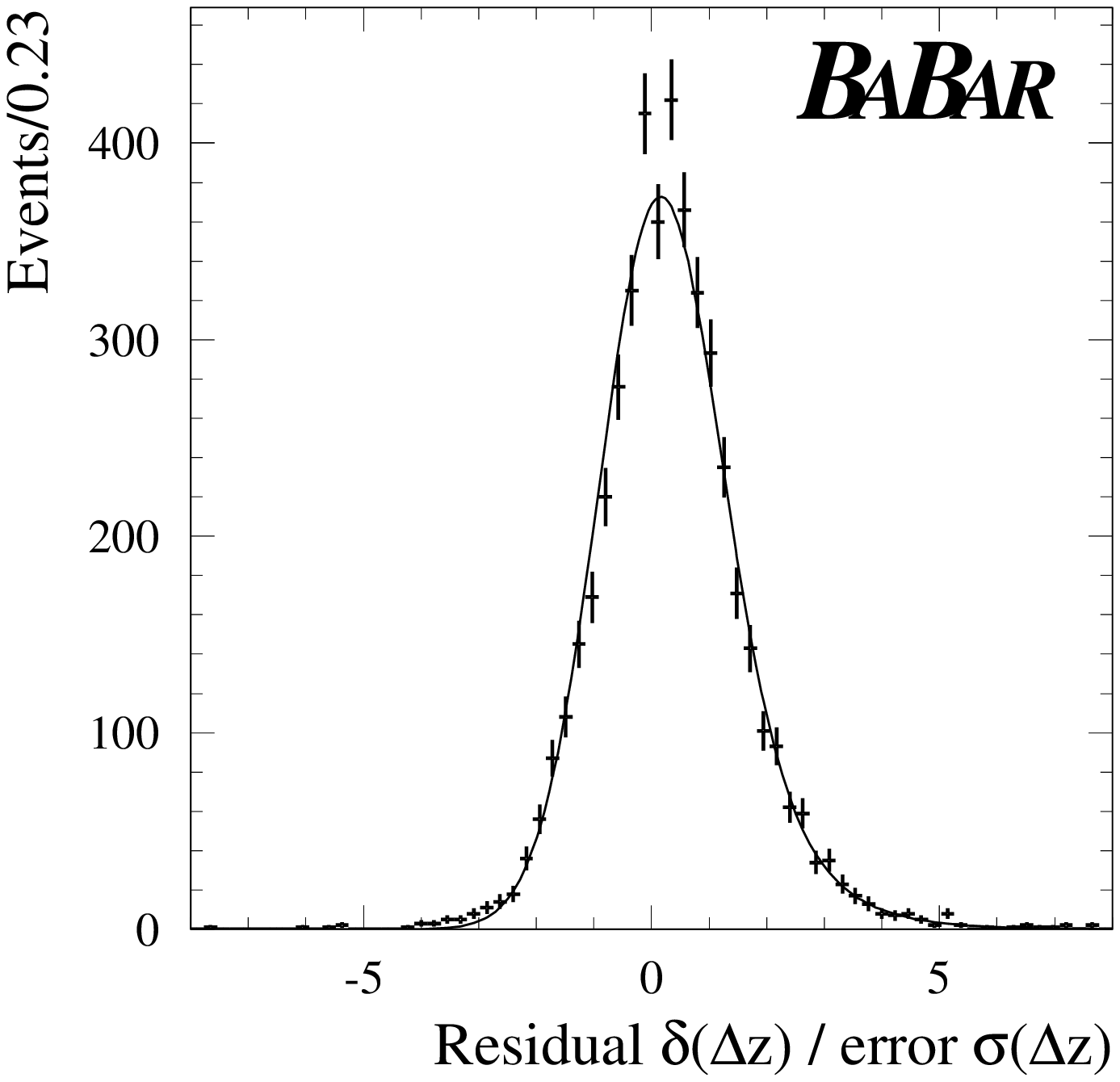,height=8cm,width=8cm}} 
%&
%\mbox{\psfig{file=Figures/fitparam2.eps,height=5cm,width=5cm}} 
%\end{tabular}
%\vskip-4cm~~~~~~~~~~~~~(a)~~~~~~~~~~~~~~~~~~~~~~~~~~~~~~~~~~~~~~~~~(b)~~~~~~~~~~~~~~~~~~~~~~~~~~~~~~~(c)\\
%%\vskip7cm~~~~~~~~~~~~~~~~(c)
%\vskip3cm
\end{center}
\caption{\label{fig:diffparams}
Fit of the $\Delta z$ pull distribution on a sample
of 5000 $B^0 \to D^- \pi^+$, $D^- \to K^+ \pi^- \pi^-$ 
Monte Carlo simulated events, with a function consisting of
a Gaussian centered 
at zero and the same Gaussian convoluted with an exponential 
($G + G \otimes E$ parametrization). 
}
\end{figure}

A common {$\cal R$}~function can be used for all modes.
The shape of the resolution function
can be fit with several parametrizations. Results for the pull
representation (which is used in the likelihood function of the
lifetime fit) are summarized in Table~\ref{tab:diffparams}.
It can be seen that a single Gaussian is too simple whereas a
two-Gaussian line shape $G+G$ fits adequately. A fair agreement is
also obtained with a function of the form $G+G\otimes E$, the sum of an
unbiased Gaussian $G$ and of the convolution of the same Gaussian with a
decaying exponential $E$, as can be seen in Fig.~\ref{fig:diffparams}.

\begin{table}[htbp]
\begin{center}
\caption{
\label{tab:vtxtagmulti}
Parameters of the $G+G\otimes E$ resolution function $\cal R$
in the pull representation for a few \Bz\ and \Bu\ decay modes. 
The parameter $g$ is
the fraction of events in the pure Gaussian, $s$ is the standard deviation
of the Gaussian function and $\tau_{\rm r}$ is the decay constant of the
exponential. A single $\cal R$ function for all modes is used in the
lifetime fit.
} 
\begin{tabular}[t]{|l|ccc|}
\hline
                        & $g$ & $s$ & $\tau_{\rm r}$\\
\hline \hline
$\Bz \to D^- \pip$     & $0.637 \pm 0.031$ & $1.019 \pm 0.017$ & $0.914 \pm 0.062$\\
$\Bz \to D^{*-} \pip$  & $0.666 \pm 0.017$ & $1.047 \pm 0.010$ & $1.024 \pm 0.041$\\
$\Bu \to \jpsi \Kp$    & $0.678 \pm 0.015$ & $1.033 \pm 0.008$ & $0.956 \pm 0.035$\\
$\Bu \to \Dzb \pip$    & $0.644 \pm 0.020$ & $1.008 \pm 0.010$ & $0.833 \pm 0.037$\\
\hline
\end{tabular}
\end{center}
\end{table}

\begin{table}[htbp]
\begin{center}
\caption{\label{tab:diffparams}
Values of the parameters for the different functional forms used to
fit the $\Delta z$ resolution function.}
\begin{tabular}[t]{|l|l|l|}
\hline
Parametrization				& Parameters 	& $\chi^2$/ndof  \\
\hline
\hline
Single Gaussian ($G$)                 & mean $=$ 0.273 $\pm$ 0.017 &\\
                                        & width $=$ 1.148 $\pm$ 0.015
                                        &  135.3/51\\
\hline
%\hline
Gaussian centered at zero plus           & $g$ $=$ 0.643 $\pm$ 0.029 &\\
the same Gaussian convoluted with an    & $s$ $=$ 1.012 $\pm$ 0.0164 &\\
exponential ($G+G \otimes E$)         & $\tau_{\rm r}$ $=$ 0.936
                                          $\pm$ 0.062 &  62.8/50\\
\hline
%\hline
Two Gaussians ($G+G$)                 & $f$   $=$ 0.808 $\pm$ 0.047 &\\
                                        & $s_1$ $=$ 0.998 $\pm$ 0.032 &\\
                                        & $b_1$ $=$ 0.204 $\pm$ 0.025 &\\
                                        & $s_2$ $=$ 1.931 $\pm$ 0.143 &\\
                                        & $b_2$ $=$ 0.819 $\pm$ 0.139
                                        & 40.0/48\\
\hline
\end{tabular}
\end{center}
\end{table}

%\begin{figure}[htbp]
%\begin{tabular}[t]{cc}
%\mbox{\psfig{file=Figures/multidz.eps,height=5cm,width=8cm}} &
%\mbox{\psfig{file=Figures/multidzpull.eps,height=5cm,width=8cm}}
%\end{tabular}
%\caption{
%\label{fig:vtxtagmulti}
%$\Delta z$ Residual (left) and pull (right) for different decay chains
%of the fully reconstructed \B. Black: $\Bub \to \jpsi K^-$, red: $\Bub
%\to D^0 \pi^-$, green: $\Bzb \to D^+ \pi^-$ and blue: $\Bzb \to D^{*+} \pi^-$.
%}
%\end{figure}

In a higher statistics Monte Carlo study, tails become apparent
which cannot be reproduced by any of the 
parametrizations considered above. These
``outlier'' events in the tail are bound to bias the lifetime
measurement if they are not
accounted for in the fit. Some are due to long lived strange particles
forming a vertex with, e.g., the $B_{\rm opp}$ pseudotrack which
is robust enough  to survive the iterative vertex finding algorithm. 
To eliminate outliers, a requirement of a minimum of two
real tracks to form the $\B_{\rm opp}$ vertex,
$|\Delta z|< 3000$\mum\ and $\sigma(\Delta z)<400$\mum\
is imposed. There remain
$1752 \pm 46$ ($1879 \pm 45$) signal events in the neutral
(charged) \B\ lifetime samples after the selection.

The distribution of $\sigma(\Delta z)$
can be fitted to a Crystal Ball line shape 
function~\cite{bib:crystalball}. The fitted function is denoted
$\rho(\sigma)$. Use of
that spectrum with the pull representation of the resolution function
provides an elegant way of implementing the lifetime fits as explained
in Section~\ref{sec:lifetimefit}.

\begin{figure}[!hbt]
\begin{center}
\begin{tabular}{lr}
\mbox{\epsfxsize=8cm\epsffile{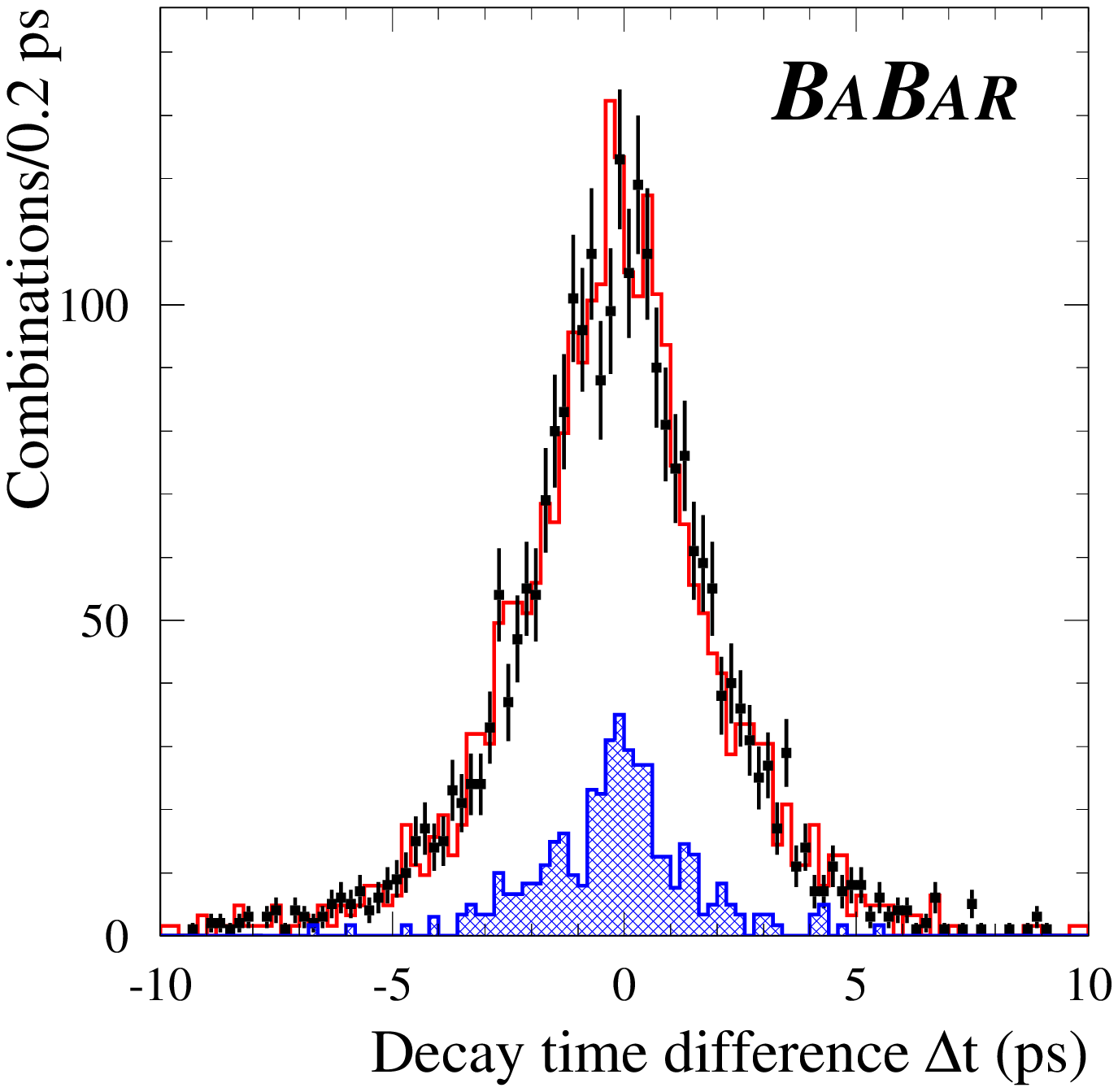}} &
\mbox{\epsfxsize=8cm\epsffile{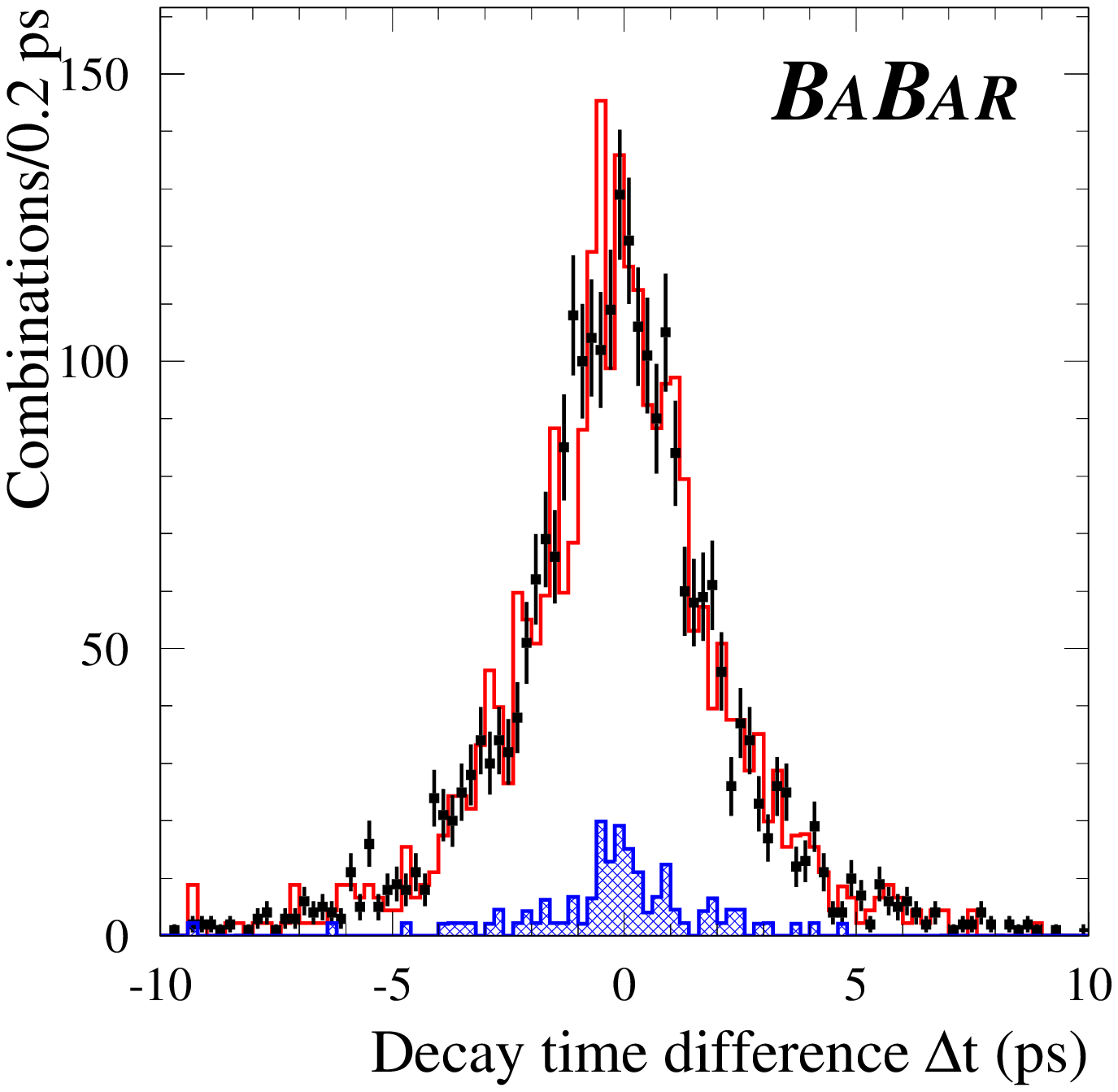}} \\
\end{tabular}
\vskip-7.7cm{\Large
  (a)}~~~~~~~~~~~~~~~~~~~~~~~~~~~~~~~~~~~~~~~~~~~~~~~~~~~~~~~~~~~~~{\Large
  (b)}~~~~~~~~~~~~~~~~~~~~~~~~~\\
\vskip7.0cm
\end{center}
\caption{
\label{fig:deltazdatamc}
Comparison of $\Delta t$ distributions in the data (points)
and \BB\ Monte Carlo events (solid line histogram) for (a) \Bz\ and
(b) \Bu. The contribution from the other type of \BB-decays in
each case and from continuum events is also shown (hatched histogram).
The simulation has been normalized to the
data statistics.}
\end{figure}

Figure~\ref{fig:deltazdatamc} shows the $\Delta z$ distributions 
in the data for candidates satisfying $\mes > 5.27$\gevcc. 
Superimposed are the
Monte Carlo simulation predictions using the central values of the
lifetimes of the \B\ mesons from Ref.~\cite{bib:PDG2000}.

%\begin{figure}
%\caption{
%\label{fig:deltazdatamc}
%Comparison of $\Delta z$ distributions in the data (points)
%and signal Monte Carlo (solid line) for $\Bub \to D^0 \pi^-$.}
%\end{figure}

\section{Lifetime fits}
\label{sec:lifetimefit}
The \B\ meson lifetimes are extracted from the $\Delta z$
distributions of the selected events with an unbinned maximum
likelihood fit. The measurements performed on each event~$i$ are
represented by three input numbers:
$(\Delta z)_i$, its error, $\sigma_i$, and $p_{{\rm sig},i}$.
All events which satisfy $5.2 < \mes < 5.3$\gevcc\ and have $\Delta E$ in
the signal region are input to the fit. $p_{{\rm sig},i}$ is the
probability for an event~$i$ to come from signal. This probability is
obtained from the independent fit of the \mes\ spectrum
using the ARGUS background function and a single Gaussian, as
described previously (see Fig.~\ref{fig:hadronicb0bch}).
The fit determines the
\B\ lifetime, $\tau_B$, the proportion of outliers, $p_{\rm out}$,
and the two sets of parameters $\theta_{\rm resol}$ and $\theta_{\rm
bkg}$ which describe the 
resolution function and the background shape, respectively.

The total probability density function (PDF) for an
event~$i$, ${\cal T}_i$,
is the sum of three terms describing the signal, the background and
the outliers,
$$ {{\cal T}_i} = \rho(\sigma_i) \times 
\left[ p_{{\rm sig},i}\ {\cal S}((\Delta z)_i, \sigma_i ; \tau_B,
\theta_{\rm resol})
+ (1-p_{{\rm sig},i} -p_{\rm out})\ {\cal B}((\Delta z)_i; \theta_{\rm bkg})
   + p_{\rm out}\ {\cal O}((\Delta z)_i) \right].$$

The signal PDF, $\cal S$, is the result of a convolution of the
theoretical $\Delta z$ distribution (two exponential wings) and the
pull representation of the $\Delta z$~resolution function.
Three or five resolution parameters, $\theta_{\rm resol}$,
(depending on which parameterization is chosen for $\cal
R$,  $G+G$ or $G + G \otimes E$) are used
to model the width and non-Gaussian tails. The parameters are free
in the lifetime fit, i.e. the values of the
parameters of the resolution function are extracted from the data themselves.
Reducing the number of 
parameters modeling $\cal R$, which are unavoidably correlated to $\tau_{B}$, 
has the effect of improving the statistical precision on the
lifetime. The parameterization needs to be flexible enough to
reproduce all features of the resolution function, however.
The optimal choice depends on the statistics available.

The events in the substituted mass sideband 
%($5.20 < \mes < 5.26$~\gevcc\ and
%$\Delta E$ in the signal region) 
carry most information on the
background $\Delta z$~distribution.
A satisfactory description of this distribution is given  by a function 
adding a single Gaussian and two independent 
exponential tails for the negative and positive $\Delta z$ ranges.
This parameterization uses six free parameters $\theta_{\rm bkg}$: the
width and mean of the Gaussian, the ``lifetimes'' of the two
exponentials and the fraction of events in each of the exponentials.
Such a shape is flexible enough to account for the combinatorial
background with continuum and \BB\ contributions.

The outlier PDF, $\cal O$, is a single wide Gaussian with a fixed width of 
2500\mum\ and a fixed mean of zero.

A full Monte
Carlo sample with 5000 events in the channel  $B^0 \to D^- \pi^+$, 
$D^- \to K^+ \pi^- \pi^-$ was used to compare different 
parameterizations of the resolution function (see
Table~\ref{tab:diffparams}) which were injected into signal-only 
lifetime fits. The $G + G$ model of the resolution function
gives an unbiased result but the
statistical error on the lifetime is increased by more than 50\%
compared to the case of a fixed $\cal R$. The single Gaussian model
is both biased (overestimation of $\simeq$ 1.5\%) and gives a 30\%
increase of the statistical error. The $G + G \otimes E$ is
intermediate with a 0.3\% overestimation of $\tau_B$ and the smallest
increase of the statistical error (10\%). With the present statistics,
the $G + G \otimes E$ model of $\cal R$ lead to the smallest overall
error. With high statistics, the $G + G$ model will become the
optimum method.

\begin{figure}[p]
\begin{center}\mbox{\psfig{file=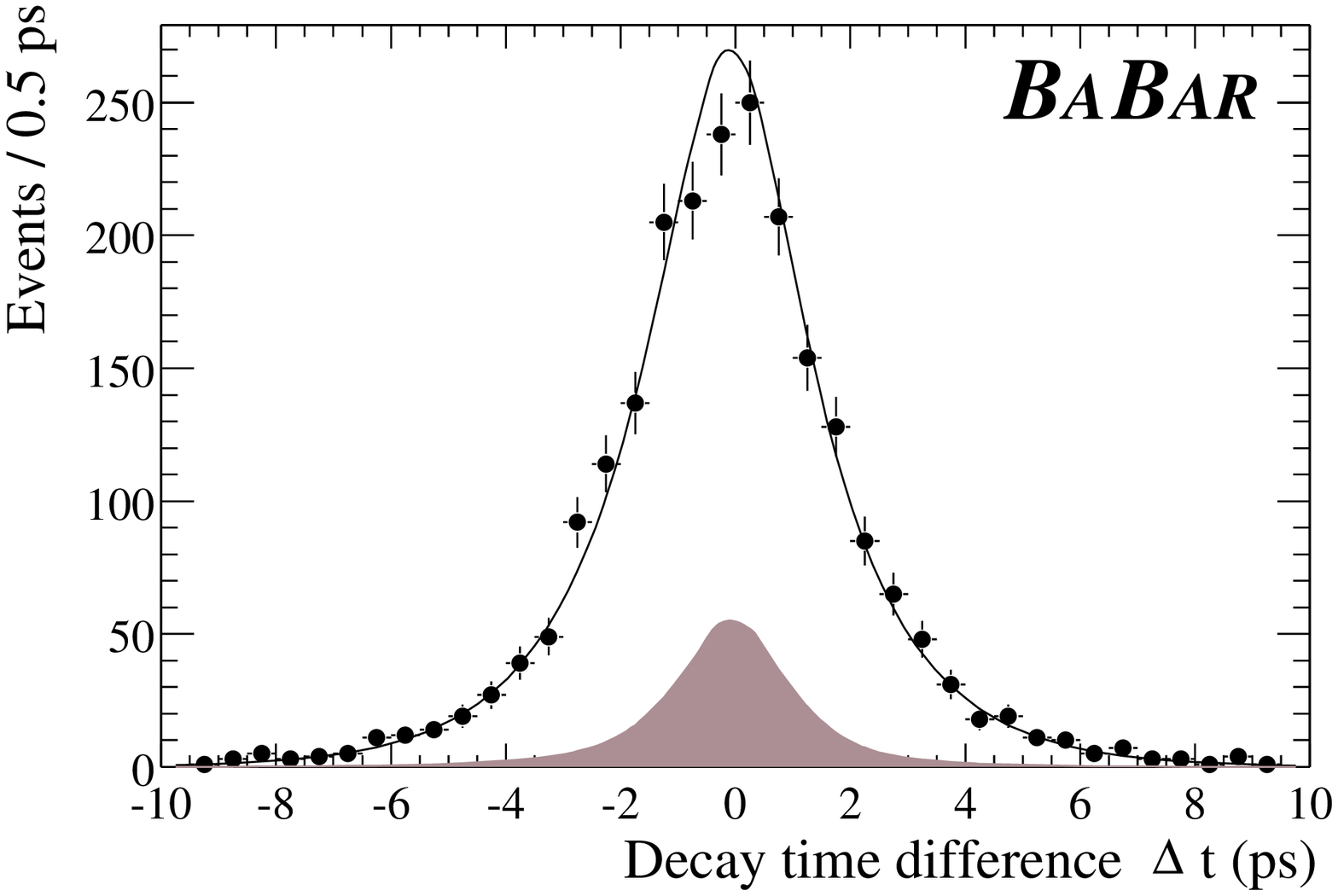,height=8cm,width=12cm}}
\end{center}
\caption{
\label{fig:FitB0}
$\Delta t$ distribution for $\Bz$/$\Bzb$~candidates in the signal
region. The result of the lifetime fit is superimposed.
The background is shown by the hatched distribution.
}
\end{figure}

\begin{figure}[htbp]
\begin{center}\mbox{\psfig{file=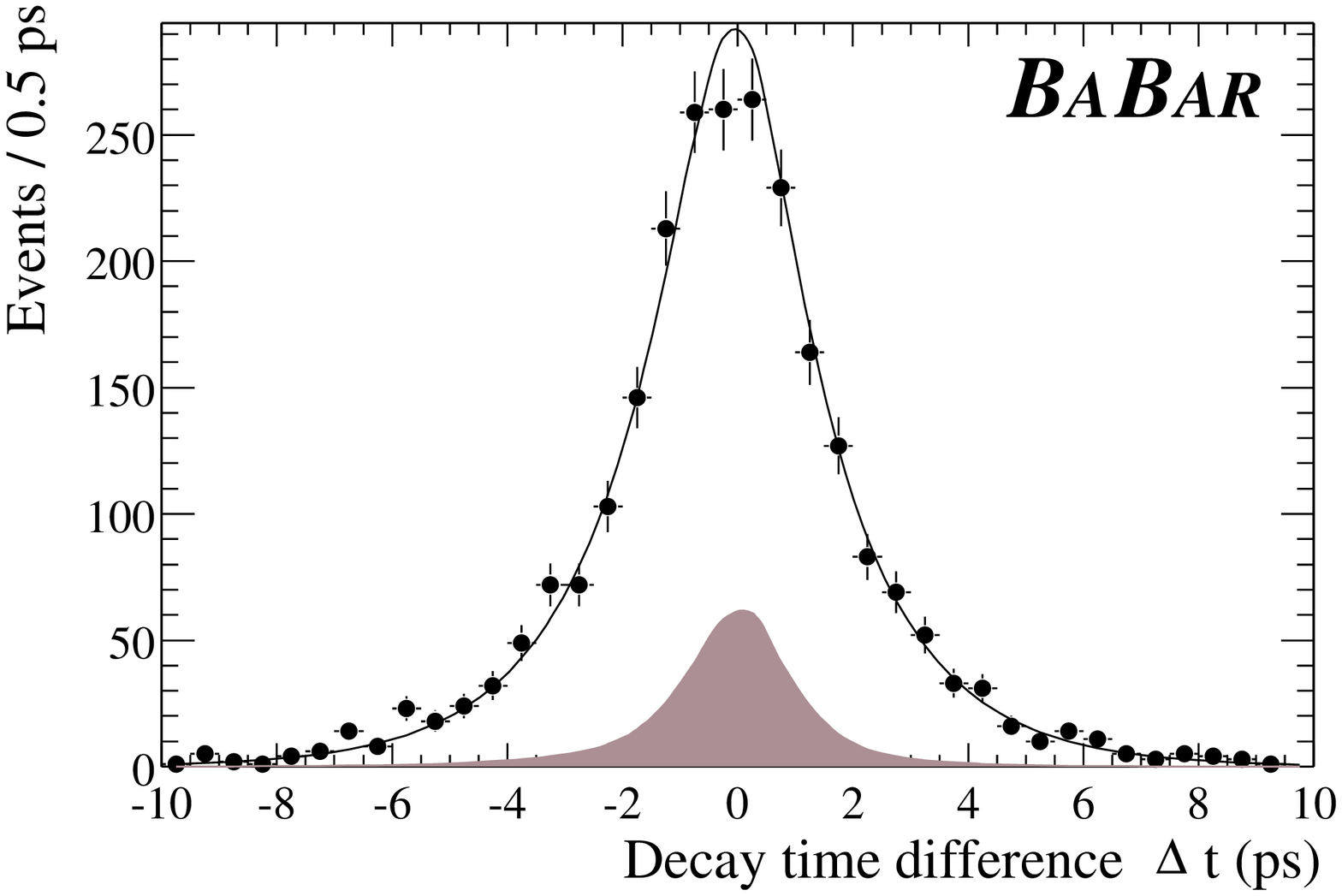,height=8cm,width=12cm}}
\end{center}
\caption{
\label{fig:FitChB}
$\Delta z$ distribution for $\Bpm$~candidates in the signal
region. The result of the lifetime fit is superimposed.
The background is shown by the hatched distribution.
}
\end{figure}

To determine the \B\ lifetimes, a combined fit to the $\Delta
z$~distributions of the \Bz\ and \Bu\ samples is performed. 
The two $\Delta z$~distributions are not combined but are fitted
simultaneously,  with different sets of parameters (one per
charge) to describe the lifetime, the background 
and the outliers. Since the resolution
functions for neutral and charged \B\ mesons are compatible, a single 
set of  parameters is used to describe
the resolution function for both samples. 
The result of the fit is
\begin{eqnarray*}
\tau_{\Bz} &=& 1.512\pm 0.052\ \ps, \\
\tau_{\Bu} &=& 1.608\pm 0.049\ \ps,
\end{eqnarray*}
where the errors are statistical only. The lifetime fit results are
superimposed upon the \Bz\ and \Bu\ $\Delta z$ distributions from the
\B~reconstruction signal region
in Figs.~\ref{fig:FitB0} and \ref{fig:FitChB}.
The lifetime ratio is determined with a similar combined fit, this
time  to the \Bz\ lifetime and the ratio, which results in a value of
$$\tau_{\Bu}/\tau_{\Bz}=1.065 \pm 0.044,$$
where, again, the error is statistical only.
The final results after corrections
described in the section on systematic errors are applied, are quoted
in section~\ref{sec:conclusions}. 

%13c13
%< to come from the signal. It is obtained from an independent fit
%---
%> to come from the signal. It is obtained from an independent fit
%32,38c32,40
%< This procedure uses the data themselves to extract the values of the
%< parameters of the resolution function.\par
%< We use events in the substituted mass sideband ($5.2 %< M_{ES} %< 5.26
%< \gevc$ and $\Delta E$ in the signal region) to study the
%< background $\Delta z$~distribution.
%< A satisfactory description of this distribution is given  by a function 
%< adding a single Gaussian and two independent 
%---
%> Note that, in this procedure, the resolution
%> function is extracted from the data and does not rely on any Monte 
%> Carlo.\par
%> The background PDF $\cal B$ parameterization was determined by 
%> studies of the $\Delta z$ distribution of events whose 
%> $M_{ES}$ falls in the sideband between 5.2 and 5.25~\gevc , 
%> and having $\Delta E$ in the signal region. 
%> A satisfactory description is given  by a function 
%> adding a single Gaussian and two independent 
%44c46,50
%< A full Monte
%---
%> The optimization of the lifetime fitting procedure is the result of
%> extensive Monte Carlo tests. The numerical soundness of the fit
%> implementation (input value reproduced with no bias, 
%> standard deviation of pull distribution equal to one) 
%> was verified on series of 500 toy Monte Carlo samples. A full Monte

\section{Systematic uncertainties}
\label{sec:systematics}
The systematic errors  associated with event selection, $\Delta z$ measurement
and lifetime fitting are discussed in turn.

It has been verified, using the \Bz\ and \Bu\ Monte Carlo samples,
that the event selection
does not distort the generated $\Delta z$ distribution so that the
lifetime measured from these samples is consistent with the input value.
These checks are done
with a limited number of simulated events which results in a statistical
error on the lifetimes extracted from the fit. This Monte Carlo
statistical error of 1.0\% for \Bz\ and 0.9\% for \Bu\ is taken as
a systematic error on the lifetime measurements,
labeled as ``MC statistics'' in Table~\ref{tab:listsystematics}.

The $\Delta z$ resolution function $\cal R$ is a key ingredient of the
lifetime measurements. The parameters describing $\cal R$ are
results of the fit. Since they are correlated with the lifetime,
part of the systematic error which comes from their uncertainty is
transferred into the statistical error on the lifetime. This effect is
estimated to be about 2\%. For the 
systematic error associated with the resolution function modeling,
two contributions are quoted. One is 
the difference between lifetimes obtained with fits using the $G+G$ and 
$G + G \otimes E$ parametrizations, a 0.3\% effect. The other,
at the level of 0.3\% as well, is an overestimation of the distortion which
may result from the simplifying assumption that a unique resolution
function is suitable for all the modes.

The sensitivity of the lifetime measurement to quantities which enter into
the determination of the $\B_{\rm opp}$ vertex, namely the beam spot position
and vertical size, and the $\B_{\rm rec}$ momentum vector has been
found to be negligible.

The fit accounts for outlier events via a wide Gaussian function with
fixed mean and width. Increasing the width from 2500 to 4000\mum\ gives
a change of 1.1\% (1.3\%) to the \Bz\ (\Bu)
lifetimes. A change of 1000\mum\ in the mean gives an effect smaller
by a factor three.

The effects of the detector geometry and alignment have been studied.
These are smaller than the effects due to the charm
lifetimes, and both parameterizations of the resolution function are
flexible enough to accommodate small changes in the detector
response. The associated systematic errors are included in those
given for the resolution function model.

An overestimation of the uncertainty on the length scale measured
along $z$ has been obtained in considering the mechanical integrity of
the SVT. A systematic error of 1\% is assumed. Actual measurements of
the length of the beam pipe with tracks are in progress and
should considerably reduce this contribution to the error.

Systematic effects due to the boost enter due to the approximation of
using the \FourS\ boost for both \B\ mesons in an event, 
as well as measurement error on the boost itself. The effect of the
approximation is estimated to be a 0.4\% shift and the measurement error from
two-prong events is 0.3\%. The central values for the lifetimes
obtained from the fit are corrected for this shift and a conservative
error of 0.4\% is assigned. Adding the two contributions in quadrature, the
systematic error due to the boost effects is 0.5\%.

The systematic error due to the signal probability, $p_{\rm sig}$, is
obtained by propagating the one standard deviation errors on the 
parameters of the line shape function to the lifetime. A 0.2\%
uncertainty is obtained.

The uncertainty in modeling the background is estimated by comparing
the Monte Carlo distributions for the background in the signal region with
the \mes\ sideband, and also comparing \mes\ sidebands in data and Monte Carlo
simulation.
The disagreement is more pronounced for \Bu\ mesons (1.1\%) than for
\Bz\ mesons (0.4\%).

The stability of the results has been checked by splitting the data
sample into subsamples of different $\B_{\rm rec}$~modes and different
times of data taking. A summary of all systematic errors as well as
their sum in quadrature is given in Table~\ref{tab:listsystematics}. 

\begin{table}[htbp]
\begin{center}
\small
\caption{
\label{tab:listsystematics}
Summary of the systematic errors for the \B\ lifetime measurements.} 
\begin{tabular}[t]{|l|r|r|r|l|}
\hline
systematic effect &  $\delta(\tau_{\Bz})$ &
$\delta(\tau_{\Bu})$ & $\delta(\tau_{\Bu}/\tau_{\Bz})$ & comment \\
 &(\fs )&(\fs )& & \\
\hline\hline
MC statistics   	                &  16  &  14 & 0.014 &\\
Parameterisation of resolution function $\cal R$ &   5  &   5 & 0.001 & some
included in the \\
					&      &     &  & statistical error, see text\\
One single          resolution function &   5  &   5 & 0.008&\\
Beamspot, $p_{B_{\rm rec}}$			&   -  &   - & - &\\
$\Delta z$ outliers              	&  16  &  20 & 0.005 &\\
Geometry and alignment			&   -  &   - & - &included in $\cal R$\\
$z$ scale                               &  15  &  16 & - & cancels in ratio\\
Boost             	                &   6  &   6 & - & cancels in ratio\\
Signal probability                      &   3  &   2 & 0.005 &\\
Background modeling                    &   5  &  17 & 0.011 &\\
\hline\hline
Total in quadrature                     &  29   & 35 & 0.021 &\\
\hline
\end{tabular}
\end{center}
\end{table}

\section{Summary}
\label{sec:conclusions}
The preliminary results for the \B\ meson lifetimes are
\begin{eqnarray*}
\tau_{\Bz} &=& 1.506 \pm 0.052\ {\rm (stat)}\ \pm 0.029\ {\rm (syst)}\ \ps, \\ 
\tau_{\Bu} &=& 1.602 \pm 0.049\ {\rm (stat)}\ \pm 0.035\ {\rm (syst)}\ \ps
\end{eqnarray*}
and for their ratio is
$$\tau_{\Bu}/\tau_{\Bz} = 1.065 \pm 0.044\ {\rm (stat)}\ \pm 0.021\ {\rm (syst)}.$$
These results are consistent with previous \B\ lifetime
measurements~\cite{bib:PDG2000} and competitive with the most precise ones.
The statistical errors dominate the uncertainties, while
the systematic uncertainty
on the lifetime ratio is smaller than the overall error of the world
average for this quantity. There are ways to reduce some
contributions; however, to reach a 1\% total error on the
\B\ lifetime ratio remains a challenge.
%However, the results described above
%were obtained by the \babar\ experiment after only one year of
%running; they are bound to improve. 

\section{Acknowledgments}
\label{sec:Acknowledgments}
%% Standard acknowledgments paragraph; must always be included.
We are grateful for the contributions of our \pep2\ colleagues in
achieving the excellent luminosity and machine conditions
that have made this work possible.
We acknowledge support from the
Natural Sciences and Engineering Research Council (Canada),
Institute of High Energy Physics (China),
Commissariat \`a l'Energie Atomique and
Institut National de Physique Nucl\'eaire et de Physique des Particules
(France),
Bundesministerium f\"ur Bildung und Forschung
(Germany),
Istituto Nazionale di Fisica Nucleare (Italy),
The Research Council of Norway,
Ministry of Science and Technology of the Russian Federation,
Particle Physics and Astronomy Research Council (United Kingdom), the
Department of Energy (US),
and the National Science Foundation (US). In addition, individual support 
has been received from the Swiss 
National Foundation, the A. P. Sloan Foundation, the Research Corporation,
and the Alexander von Humboldt Foundation.
The visiting groups wish to thank 
SLAC for the support and kind hospitality
extended to them.     

%\section*{References}


\begin{thebibliography}{99} 

%\bibitem{bib:lepwg} LEP \B~lifetimes working group. Homepage:\\
%http://claires.home.cern.ch/claires/lepblife.html~.

\bibitem{bib:Bigi} I.\ I.\ Bigi, Nuovo Cim.~{\bf 109~A}, 713 (1996).
%``Lifetimes of heavy flavor hadrons:
%whence and whither~?'',

\bibitem{bib:Neubert} M.\ Neubert and C.\ T.\ Sachrajda,
Nucl. Phys. B {\bf 483}, 339 (1997).
%``Spectator effects
%in inclusive decays of beauty hadrons''

\bibitem{bib:PDG2000} Particle Data Group, D.\ E.\ Groom {\em et al.},
\epjc{15}, 1 (2000).

\bibitem{bib:babar}
\babar\ Collaboration, B.\ Aubert {\em et al.}, ``The first year of the
\babar\  experiment at \pep2'', \babar-CONF-00/17, submitted to the XXX$^{th}$
International Conference on High Energy Physics, Osaka, Japan.

\bibitem{bib:babbook113} P. F. Harrison and H. R. Quinn, eds., ``The
\babar\ physics book'', SLAC-R-405 (1998), section 11.3.

%\bibitem{bib:joe} DIRC - The particle identification system for BABAR,
%contribution to ICHEP2000 in parallel session PA12c presented by
%J.~Schwiening for the \babar\ Collaboration.

\bibitem{bib:BabarPub0005} 
\babar\ Collaboration, B.\ Aubert {\em et al.},
``Exclusive \B~decays to charmonium final states'',
\babar-CONF-00/05, submitted to the XXX$^{th}$
International Conference on High Energy Physics, Osaka, Japan.

%\bibitem{bib:brownchep} Kalman filter, talk by D.N. Brown at CHEP-2000.
%\bibitem{bib:btodbranchingratios} conf 00-06 paper.

\bibitem{bib:argusfunction} ARGUS collaboration, H.\ Albrecht {\em et al.}, 
Z~Phys.~{\bf C48}, 543 (1990).

\bibitem{bib:crystalball} T. Skwarnicki, ``A Study of the Radiative
Cascade Transitions between the Upsilon-Prime and Upsilon Resonances'',
DESY~F31-86-02 (thesis, unpublished) (1986).
\end{thebibliography}
\end{document}